\documentclass[aps,superscriptaddress,preprintnumbers,showpacs]{revtex4}

\usepackage{txfonts}

\usepackage{amssymb}

\usepackage{indentfirst}

\usepackage{epsfig}

\usepackage{epsf}

\usepackage{graphicx}

\newcommand{\be}{\begin{equation}}
\newcommand{\ee}{\end{equation}}
\newcommand{\ba}{\begin{eqnarray}}
\newcommand{\ea}{\end{eqnarray}}

\begin{document}

\newcommand*{\LZ}{School of Physical Science and Technology, Lanzhou University, Lanzhou 730000, China}
\affiliation{\LZ}
\newcommand*{\UC}{Centro de F\'{i}sica Computacional, Departamento de
F\'{i}sica, Universidade de Coimbra, P-3004-516 Coimbra,
Portugal}\affiliation{\UC}
\newcommand*{\RU}{Institute for Theoretical Physics, Regensburg
University, D-93040 Regensburg, Germany}\affiliation{\RU}
\newcommand*{\NKU}{Physics Department, Nankai University, 300071, Tianjin,
China}\affiliation{\NKU}


\title{The puzzle of excessive non-$D\bar D$ component of the inclusive $\psi(3770)$
decay and the long-distant contribution}

\author{Xiang Liu}\email{liuxiang@teor.fis.uc.pt}\affiliation{\LZ}\affiliation{\UC}
\author{Bo Zhang}\affiliation{\RU}
\author{Xue-Qian Li}\email{lixq@nankai.edu.cn}\affiliation{\NKU}

\date{\today}

\begin{abstract}
In this letter we suggest that the obvious discrepancy between
theoretical prediction on the $\mathrm{non}-D\bar D$ decays of
$\psi(3770)$ and data is to be alleviated by taking final state
interaction (FSI) into account. By assuming that $\psi(3770)$
overwhelmingly dissociates into $D\bar D$, then the final state
interaction induces a secondary process, we calculate the
branching ratios of $\psi(3770)\to D\bar D\to
J/\psi\eta,\,\rho\pi,\,\omega\eta,\,K^*K$. Our results show that
the branching ratio of $\psi(3770)\to \mathrm{non}-D\bar{D}$ can
reach up to
$\mathcal{B}_{\mathrm{non}-D\bar{D}}^{FSI}=(0.2\sim1.1)\%$ while
typical parameters $I=0.4$ GeV$^{-2}$ and $\alpha=0.8\sim 1.3$ are
adopted. This indicates that the FSI is obviously non-negligible.

\end{abstract}

\maketitle


Obviously, physics about Charmonium is still not a closed book
yet, instead, this field is full with challenges and opportunities
\cite{Li:2008ey}. Due to developments and improvements of facility
and technique, the detection precision is greatly enhanced in the
past decade, as a direct consequence new puzzles have continually
emerged. Indeed, some old puzzles have been understood now, but a
number of them remain unsolved yet. Theorists are endeavoring to
look for solutions. The general strategy is that first, one can
fumble solutions in the framework of present theory i.e. QCD and
see if we miss something in our calculations, then if all
possibilities are exhausted one needs to invoke new physics beyond
the standard
model. 
In this work, we follow the first strategy to explain excessive
non-$D\bar D$ component of inclusive $\psi(3770)$ decay, which is
conducted in a series of experiments by the BES collaboration
\cite{Bai:2003hv,Ablikim:2006zq,Ablikim:2006aj,Ablikim:2007zz,Ablikim:2008zz}
in the past three years.

As a well measured charmonium state, $\psi(3770)$ generally is
considered as a mixture of $2^{3}{S}_1$ and $1^{3}D_{1}$ states
\cite{Ding:1991vu,Rosner:2001nm}. Since 3770 MeV is a bit above
the threshold of $D\bar D$ production, such a bound state may
dissolve into open charms which eventually hadronize into $D\bar
D$. Therefore, before observing sizable non-$D\bar{D}$ decay
rates, $\psi(3770)$ was supposed to dominantly decay into
$D\bar{D}$, including $D^0\bar{D}^0$ and $D^+D^-$. There could be
some possible non-$D\bar{D}$ modes \cite{
Adam:2005mr,Coan:2005ps,Huang:2005fx,Adams:2005ks,CroninHennessy:2006su,Briere:2006ff},
especially the hidden charm decay modes, such as $J/\psi\pi\pi$
and $J/\psi\eta$ with $\mathcal{B}[\psi(3770)\to
J/\psi\pi^+\pi^-]=(1.93\pm0.28)\times10^{-3}$,
$\mathcal{B}[\psi(3770)\to
J/\psi\pi^0\pi^0]=(8.0\pm3.0)\times10^{-4}$ and
$\mathcal{B}[\psi(3770)\to J/\psi\eta]=(9\pm 4)\times10^{-4}$
respectively \cite{Amsler:2008zz}, 
and E1 radiative decays $\gamma \chi_{cJ}$ with decay widths
$172\pm 30$ keV, $70\pm17$ keV and $<21$ keV for $J=0,\,1,\,2$
respectively \cite{Coan:2005ps,Briere:2006ff}. The sum of all the
branching ratios of these hidden charm decay modes is less than
2\%, so all these measurements support the allegation that
$\psi(3770)$ overwhelmingly decays into $D\bar D$.

However, the BES collaboration investigated the inclusive decays
of $\psi(3770)$ and found that the branching ratio of
$\psi(3770)\to D\bar D$ is about $(85\pm 5)\%$
\cite{Ablikim:2006aj,Ablikim:2006zq}.  This is later verified by
the measurements of non-$D\bar{D}$ inclusive processes with the
branching fraction $\mathcal{B}[\psi(3770)\to
\mathrm{non}-D\bar{D}]=(13.4\pm5.0\pm 3.6)\%$
\cite{Ablikim:2007zz} and $\mathcal{B}[\psi(3770)\to
\mathrm{non}-D\bar{D}]=(15.1\pm5.6\pm 1.8)\%$
\cite{Ablikim:2008zz} respectively by adopting two different
methods. The CLEO measurements indicate
$\sigma(e^+e^-\to\psi(3770)\to
\mathrm{hadrons})=(6.38\pm0.08^{+0.41}_{-0.30})$ nb
\cite{Besson:2005hm} and $\sigma(e^+e^-\to\psi(3770)\to
D\bar{D})=(6.57\pm0.04\pm 0.01)$ nb \cite{:2007zt}, which together
make a $\mathcal{B}(\psi(3770)\to D\bar{D})=(103.0\pm
1.4_{-6.8}^{+5.1})\%$. Notice that the error on the high side is
about $6.8\%$ \footnote{We thank Dr. Brian Heltsley and Dr. Hajime
Muramatsu for informing us of  the results of CLEO and indicating
some details about the error estimate, we then employ their
measured values in  our numerical computations, and make a
comparison with the results based on the BES data.}, by this error
tolerance, there could be a large $(10\sim 15)\%$ fraction of
$\psi(3770)\to \mathrm{non}-D\bar D$ decays. The CLEO and BES
results are inconsistent at $> 2\sigma$  level, and we would
employ both of them as inputs to our numerical computations and an
obvious difference is explicitly noticed.

Kuang and Yan \cite{Kuang:1989ub} calculated the $\psi(3770)\to
J/\psi\pi\pi$ using the QCD multi-expansion, which properly deals
with the emission of light hadrons during heavy quarkonia
transitions (for a review see an enlightening paper
\cite{Kuang:2006me}). Their prediction is consistent with the
exclusive measurement on hidden charm decays of $\psi(3770)$. It
is generally concurred that, the measurements on the well measured
channels $J/\psi\pi\pi$, $J/\psi\eta$ and $\gamma \chi_{cJ}$ are
consistent with present theoretical predictions. Thus to
understand the experimental results, one should find  where
$\psi(3770)$ goes besides $J/\psi\pi\pi$, $J/\psi\eta$ and $\gamma
\chi_{cJ}$. Recently He, Fan and Chao \cite{He:2008xb} introduced
the color-octet mechanism and calculated the $\psi(3770)\to
\mathrm{light\, hadrons}$ in the framework of NRQCD by considering
next to leading order contribution. The calculation result shows
that $\Gamma[\psi(3770)\to \mathrm{light\, hadrons}]$ is
$467^{-187}_{+338}$ keV. If combing radiative decay contribution
with that of $\psi(3770)\to \mathrm{Light\, Hadrons}$, the
branching ratio of the non-$D\bar{D}$ of $\psi(3770)$ is about
$5\%$ \cite{He:2008xb}, which is still three times smaller than
$15\%$ non-$D\bar D$ branching ratio measured by the experiment.

Instead, Voloshin suggested, $\psi(3770)$ is not a pure $c\bar{c}$
state. There exists a sizable four quark component $(u\bar{u}\pm
d\bar{d})c\bar c$  and the fraction is about $\mathcal{O}(10\%)$
in $\psi(3770)$, which results in a measurable rate of
$\psi(3770)\to \pi^0 J/\psi,\,\eta J/\psi$ \cite{Voloshin:2005sd}.

Generally, one can categorize the strong decay modes of $\psi(3770)$
into three types: open charm decay ($D\bar{D}$), hidden charm decay
$J/\psi X$ ($X=\mathrm{light\, mesons}$) and the decay into light
hadrons (L-H decay). One can be more confident that the rates of
hidden charm decays are properly evaluated in terms of the QCD
multi-expansion, and the L-H decay occurs via three-gluon emission
mechanism $c\bar{c}\to 3g$.

\begin{figure}[htb]
\begin{center}
\begin{tabular}{cccccccc}
\scalebox{0.9}{\includegraphics{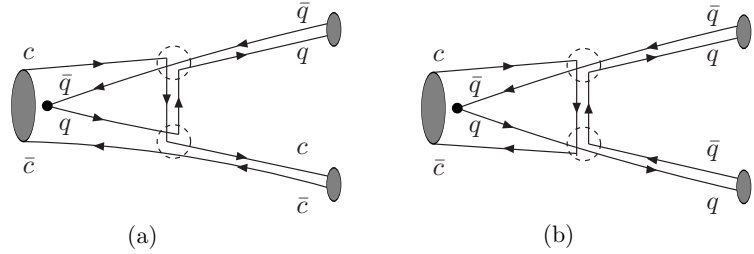}}
\end{tabular}
\end{center}
\caption{Quark-level descriptions of the hadronic loop mechanism for
the hidden charm decay (diagram (a)) and L-H decay (diagram (b)) of
$\psi(3770)$.\label{quark-level}}
\end{figure}

There is an alternative explanation to the puzzle. Twenty years
ago, Lipkin proposed that the non-$D\bar{D}$ strong decays of
$\psi(3770)$ realize via $D\bar D$ intermediate states, and
further suggested that $\psi(3770)$ does not $100\%$ decay into
$D\bar{D}$ \cite{Lipkin:1986av}. Later Achasov and Kozhevnikov
calculated the non-$D\bar{D}$ channels of $\psi(3770)$ only
considering the contribution from the imaginary part of the decay
amplitude \cite{Achasov:2005qb}. Namely such final state
interactions which are involved in the hadronic loop effects, do
contribute to both the hidden charm and L-H decays. The essential
point of the loop effect is attributed to the coupled channel
effects. A quark-level process is explicitly illustrated in the
left diagram of Fig. \ref{quark-level}. Such a mechanism should
exist in all hidden charm and L-H decays of charmonia
\cite{Liu:2006dq,Liu:2006df}. As shown in Fig. \ref{quark-level},
$\psi(3770)\to J/\psi X$ and $\psi(3770)\to \mathrm{Light\,
Hadrons}$ processes do not suffer from the Okubo-Zweig-Iizuka
(OZI) suppression. Since $\psi(3770)\to D\bar D$ takes place near
the energy threshold, one can expect that the FSI may be
significant.

In this letter, we focus on two-body hidden charm decay modes
($J/\psi\eta$) and two-body L-H decay modes ($\rho\pi$,
$\omega\eta$ and $K^*{K}$) which obviously are the main ones. Here
$K^*{K}$ denotes $K^*\bar{K}+\bar{K}^*{K}$.

\begin{center}
\begin{figure}[htb]
\begin{tabular}{c}
\scalebox{0.9}{\includegraphics{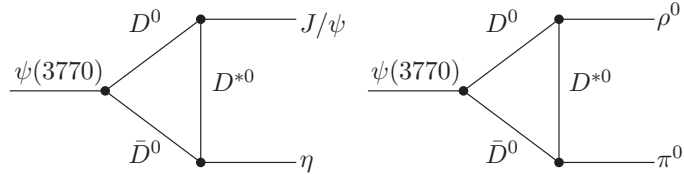}}
\end{tabular}
\caption{The hadron-level diagrams depicting hadronic loop effect
on $\psi(3770) \to D^{0}\bar{D^{0}}\to J/\psi\eta,\,\rho^0\pi^0$.
Of course they can be simply replaced by other states
$D^{(*)0}(\bar{D}^{(*)0})\to D^{(*)+}(D^{(*)-})$ with an isospin
transformation and $D^{(*)}\to\bar{D}^{(*)}$ with a charge
conjugate transformation to constitute new but similar diagrams.
By replacing relevant mesons, we can obtain the diagrams for
$\omega\eta$ and $K^*{K}$ channels. \label{hadron-level-1}}
\end{figure}
\end{center}

In order to calculate the hadronic loop effect in strong decays of
$\psi(3770)$, we consider the diagrams depicted by Fig.
\ref{hadron-level-1}, which are an alternative description in the
hadron-level language. $\psi(3770)$ first dissolves into two charmed
mesons, then by exchanging $D^{*}$ in t-channel, they turn into two
on-shell real hadrons $\mathcal{A}$ and $\mathcal{B}$. Since the
dissociation does not suffer from the OZI suppression, one can
expect it to be dominant.

One can obtain the absorptive part of the decay amplitude of
$\psi(3770)\to D+\bar{D}\to \mathcal{A+B}$
($\mathcal{AB}=J/\psi\eta$, $\rho\pi$, $\omega\eta$ and $K^*{K}$)
\begin{eqnarray}
A_{\mathcal{AB}}(M_{\psi}^2)&=&\frac{|\mathbf{p}|}{32\pi^2
M_{\psi}}\int d \Omega \mathcal{M}^*\left[\psi(3770)\to
D\bar{D}\right]\nonumber\\&&\times\mathcal{M}\left[D\bar D\to
\mathcal{AB}\right]\,\cdot\,\mathcal{F}^2\left[m_{D^{(*)}}^2,q^2\right]\,,\label{amp}
\end{eqnarray}
where
$|\mathbf{p}|=\left[\lambda(M_{\psi}^2,m_{D}^2,m_{D}^2)\right]^{1/2}/(2M_{\psi})$
is the three-momentum of the intermediate charmed mesons in the
center of mass frame of $\psi(3770)$.
$\lambda(a,b,c)=a^2+b^2+c^2-2ab-2ac-2bc$ is the K\"{a}llen
function. The form factor
$\mathcal{F}\left[m_{D^{(*)}}^2,q^2\right]$ is the key point for
the evaluation of the amplitude. One can use the monopole form
factor (FF)
\begin{eqnarray}
\mathcal{F}\left[m_{i}^2,q^2\right]=\frac{\Lambda(m_i)^2-m_i^2}{\Lambda(m_i)^2-q^2}
\end{eqnarray}
which compensates the off-shell effect of exchanged meson and
describes the structure effect of the interaction vertex. As a
free parameter, $\Lambda$ can be parameterized as
$\Lambda(m_{i})=m_{i}+\alpha\Lambda_{QCD}$ \cite{Cheng:2004ru}.
$m_{i}$ denotes the mass of exchanged meson, $\Lambda_{QCD}=220$
MeV. The range of dimensionless phenomenological parameter
$\alpha$ is around $0.8<\alpha<2.2$ \cite{Cheng:2004ru}. As a
matter of fact, there are other possible forms for
$\mathcal{F}\left[m_{D^{(*)}}^2,q^2\right]$, such as the
exponential one etc. in literature. Generally they are equivalent
somehow, as long as their asymptotic behaviors are the same.

Since the mass of $\psi(3770)$ is close to the threshold of
$D\bar{D}$ production, the dispersive part of the amplitude of
$\psi(3770)\to D+\bar{D}\to \mathcal{A+B}$ makes a large
contribution to the decay width. By unitarity, one can obtain the
dispersive part in terms of the dispersion relation. The total
decay amplitude of $\psi(3770)\to D+\bar{D}\to \mathcal{A+B}$
which includes both absorptive and dispersive parts is expressed
by \cite{Meng:2007cx,Liu:2007qi,Zhang:2007su}
\begin{eqnarray}
&&\mathcal{M}\left[\psi(3770)\to D\bar{D}\to \mathcal{A\,B}\right]
\nonumber\\&&=\frac{1}{\pi}\int_{r_{D\bar
D}^2}^{\infty}\frac{A_\mathcal{AB}(r)\,
\mathcal{R}(r)}{r-M_{\psi}^2}\,dr \,+A_{\mathcal{AB}}(M_{\psi}^2)\,,
\end{eqnarray}
where $r_{D\bar D}^2=4m_{D}^2$. After replacing $M_{\psi}^2$ in
the amplitude in eq. (\ref{amp}) with $r$, we get the amplitude
$A_{\mathcal{AB}}(r)$. The energy dependent factor
$\mathcal{R}(r)$ is defined as $
\mathcal{R}(r)=\exp\,(-\mathcal{I}\, |\mathbf{q}( r)|^2)$, which
not only reflects the $|\mathbf{q}(r)|-$dependence of the
interaction between $\psi(3770)$ and $D\bar{D}$ mesons, but also
plays the role of ultraviolet cutoff. Meanwhile, $\mathcal{R}(r)$
can be understood as the coupled channel effect summing up all the
bubbles from the charmed meson loops \cite{vanBeveren:2007cb}.
Here $|\mathbf{q}(r)|$ denotes the three momentum of $D$ meson in
the rest frame of $\psi(3770)$ with the mass $M_{\psi}\approx
\sqrt{r}$. The interaction length factor $\mathcal{I}$ is related
to the radius of the interaction by $\mathcal{I}= R/6$
\cite{Pennington:2007xr}. Pennington and Wilson indicated that
$\mathcal{I}=0.4$ GeV$^{-2}$ corresponding to $R=0.3$ fm is
favorable when studying the charmonium mass shift
\cite{Pennington:2007xr}. 

Based on the effective Lagrangian approach, one can formulate
$\mathcal{M}^*[\psi(3770)\to D\bar D]$ and $\mathcal{M}[D\bar{D}\to
\mathcal{AB}]$. The effective Lagrangians related to our calculation
are constructed by considering the chiral and heavy quark symmetries
\cite{Yan:1992gz,Wise:1992hn,Casalbuoni:1996pg}
\begin{eqnarray*}
&&\mathcal{L}_{\Psi DD}=ig_{{\Psi
\mathcal{D}{\mathcal{D}}}}\left[{\mathcal{D}_{i}}{\stackrel{\leftrightarrow}{\partial}}
 _{\mu}\mathcal{D}^{j\dag}\right]\epsilon^{\mu}\,,\\
&& \mathcal{L}_{\Psi DD^{*}}=g_{{\Psi
\mathcal{D}{\mathcal{D}^{*}}}}\epsilon^{\mu\nu\alpha\beta}\epsilon_{\mu}
\partial_{\nu}{\mathcal{D}_{i}}\partial_{\beta}{\mathcal{D}}^{*j\dag}_{\alpha}\,,
\\
&&\mathcal{L}_{D^{*}D\mathbb{P}}=-ig_{{\mathcal{D^{*}}\mathcal{D}\mathbb{P}}}
\left(\mathcal{D}^{i}\partial^{\mu}\mathbb{P}_{ij}\mathcal{D^{*}}_{\mu}^{j\dagger}
-\mathcal{D^{*}}_{\mu}^{i}\partial
^{\mu}\mathbb{P}_{ij}\mathcal{D}^{j\dagger}\right)\,,\\
&&\mathcal{L}_{D^{*}D\mathbb{V}}=
-2f_{{\mathcal{D^{*}}\mathcal{D}\mathbb{V}}}\varepsilon_{\mu\nu\alpha\beta}\left(\partial^{\mu}\mathbb{V}^{\nu}\right)
^{i}_{j}\,\Big(\mathcal{D}_{i}^{\dagger}
{\stackrel{\leftrightarrow}{\partial}}^{\alpha}\mathcal{D^{*}}^{\beta
j}-\mathcal{D^{*}}_{i}^{\beta
\dagger}{\stackrel{\leftrightarrow}{\partial}}^{\alpha}\mathcal{D}^{j}\Big)\,,
\label{lagrangian}
\end{eqnarray*}
where $\Psi$ denotes charmonium states  $J/\psi$ and
$\psi(3770)$. 
$\mathbb{P}$ and $\mathbb{V}$ are the octet pseudoscalar and nonet
vector meson matrices, respectively. The values of coupling
constants relevant to our calculation are $g_{{\psi
\mathcal{DD}}}=4.70$, $g_{{J/\psi \mathcal{DD}^*}}=4.25$
GeV$^{-1}$, $g_{{\mathcal{D}^*\mathcal{D}\mathbb{P}}}=17.31$ and
$f_{{\mathcal{D}^*\mathcal{D}\mathbb{V}}}=2.33$ GeV$^{-1}$
determined in Refs.
\cite{Liu:2006dq,Yan:1992gz,Casalbuoni:1996pg}.

For the process $\psi(3770)\to D(k_{1})+\bar{D}(k_{2})\to
J/\psi(k_3)+\eta(k_{4})$ by exchanging $D^{*0}$, one formulates
its amplitude {\small
\begin{eqnarray}
{A}_{J/\psi
\eta}&=&\mathcal{G}_{J/\psi\eta}\,\mathcal{Q}_{J/\psi\eta}\,\frac{|\mathbf{p}|}{32\pi^2
M_{\psi}}\int \mathrm{d}\Omega\left[ig_{{\psi \mathcal{DD}}}
(k_{1}-k_{2})\cdot
\epsilon_{\psi}\right]\nonumber\\&&\times\left[i\,g_{{J/\psi}\mathcal{DD}^{*}}
\varepsilon_{\kappa\xi\lambda\tau}\epsilon_{J/\psi}^{\kappa}(-i\,k_{1}^{\xi})(iq^{\tau})\right]
\left[g_{\mathcal{D}^{*}\mathcal{D}\mathbb{P}}(i\,k_{4\mu})\right]\nonumber\\
&&\times\frac{i}{q^2 -m_{D^*}^2
}\left(-g^{\mu\lambda}+\frac{q^{\mu}q^{\lambda}}{m_{D^*}^2}\right)\,\mathcal{F}^2\left[m_{D^*}^2
,q^2\right].\label{aaaa1}
\end{eqnarray}}
The absorptive amplitude of $\psi(3770)\to
D(k_{1})+\bar{D}(k_{2})\to \mathbb{P}(k_3)+\mathbb{V}(k_{4})$
reads as
\begin{eqnarray}
{A}_{\mathbb{PV}}&=&\mathcal{G}_{\mathbb{PV}}\,\mathcal{Q}_{\mathbb{PV}}
\,\frac{|\mathbf{p}|}{32\pi^2 M_{\psi}}\int
\mathrm{d}\Omega\left[i\,g_{\psi \mathcal{DD}} (k_{1}-k_{2})\cdot
\epsilon_{\psi}\right]\nonumber\\&&\times
\left[-2i\,f_{\mathcal{D}^{*}\mathcal{D}\mathbb{V}}\varepsilon_{\kappa\xi\tau\lambda}(i\,k_{3}^{\kappa})
\epsilon_{\mathbb{V}}^{\xi}(-ik_{1}^{\tau}-iq^{\tau})\right]\left[g_{\mathcal{D}^*\mathcal{D}\mathbb{P}}
(i\,k_{4\mu})\right]\nonumber\\
&&\times
\left(-g^{\mu\lambda}+\frac{q^{\mu}q^{\lambda}}{m_{D^*}^2}\right)\,\,
\frac{i}{q^2 -m_{D^*}^2
}\,\,\mathcal{F}^2\left[m_{D^*}^2,q^2\right]\,. \label{aaaa2}
\end{eqnarray}
Here the isospin factor from $\mathbb{P}$ and $\mathbb{V}$
matrices results in an extra factor $\mathcal{G}_{\mathcal{AB}}$
in the above amplitudes, which are $1/\sqrt{6}$, $1/2$, $1$ and
$1/\sqrt{12}$ for the amplitudes of $J/\psi\eta$, $\rho^0\pi^0$,
$K^{*+}K^-$ $(K^{*-}K^+,\,K^{*0}\bar{K}^0,\,\bar{K}^{*0}K^0)$,
$\omega\eta$ modes, respectively. If considering SU(3) symmetry,
the factor $\mathcal{Q}_{\mathcal{AB}}$ comes from both the
isospin transformation $D^{0(*)}\to D^{(*)\pm}$ and the charge
conjugate transformation $D^{(*)}\to\bar{D}^{(*)}$, which results
in $\mathcal{Q}_{\mathcal{AB}}=4,\, 2$ for the amplitudes of
$J/\psi\eta\,(\rho^0\pi^0,\,\omega\eta)$ and $K^{*+}K^-\,(
K^{*-}K^+,\,K^{*0}\bar{K}^0,\bar{K}^{*0}K^0)$ channels,
respectively.

\begin{figure}[htb]
\begin{tabular}{cc}
\scalebox{1.2}{\includegraphics{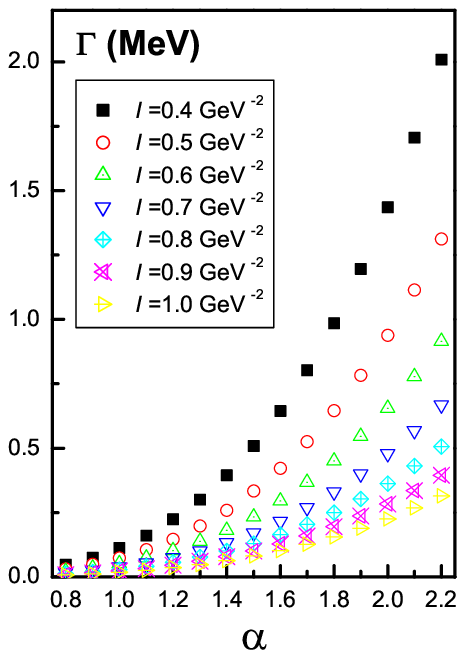}}&\scalebox{1.2}{\includegraphics{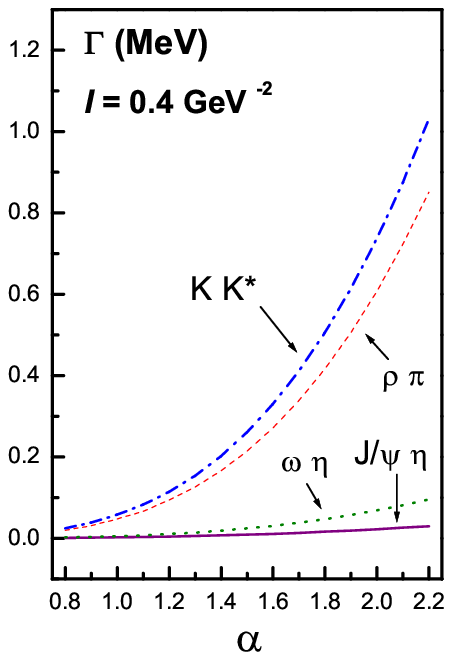}}
\end{tabular}
\caption{The dependence of decay width of sum of $\psi(3770)\to
J/\psi\eta,\,\rho\pi,\,\omega\eta,\,K^*K$ channels on $\alpha$
with several typical value $\mathcal{I}=0.4\sim 1.0$ GeV$^{-2}$.
In right-hand diagram, with $\mathcal{I}=0.4$ GeV$^{-2}$, we also
show the variation of partial decay widths of $\psi(3770)\to
J/\psi\eta,\, \rho\pi,\,\omega\eta,\,K^*K$ with $\alpha$.
\label{width}}
\end{figure}
In the left diagram of Fig. \ref{width}, we plot the dependence of
numerical result on $\alpha$ with several typical value
$\mathcal{I}=0.4\sim 1.0$ GeV$^{-2}$, which is the sum of decay
widths of $\psi(3770)\to \rho\pi,\,\omega\eta,\,K^*K$ induced by
$D\bar{D}$ intermediate states. We set $\Gamma[\psi(3370)\to
D\bar{D}\to\rho\pi]\approx 3\Gamma[\psi(3370)\to
D\bar{D}\to\rho^0\pi^0]$ and $\Gamma[\psi(3370)\to
D\bar{D}\to{K^*K}]\approx 4\Gamma[\psi(3370)\to D\bar{D}\to
K^{*+}K^{-}]$ which are determined by the SU(3) symmetry. The
dependence of decay widths of each modes $\psi(3770)\to
J/\psi\eta,\, \rho\pi,\,\omega\eta,\,K^*K$ induced by long-distant
contribution on the parameter $\alpha$ within the range of
$0.8\leq\alpha \leq 2.2$ is shown in the right diagram of Fig.
\ref{width}, where a typical value $\mathcal{I}=0.4$ GeV$^{-2}$ is
adopted.

Our numerical results indicate that the decay widths of
$\psi(3770)\to D\bar{D}\to\rho\pi,\,K^*K$ are about one order
larger than that of $\psi(3770)\to D\bar{D}\to
J/\psi\eta,\,\omega\eta$ as we set the same values of parameters
$\alpha$ and $\mathcal{I}$ for all the processes. The difference
between the widths of $\psi(3770)\to D\bar{D}\to\rho\pi,\,K^*K$
and that of $\psi(3770)\to D\bar{D}\to J/\psi\eta,\,\omega\eta$ is
due to the phase space, factors $\mathcal{G}_{\mathcal{AB}}$ and
$\mathcal{Q}_{\mathcal{AB}}$. Whereas the amplitudes for
$\rho\pi,\,K^*K$ are comparable, and they are the main
$\mathrm{non}-D\bar D$ decay channels obviously.

\begin{figure}[htb]
\scalebox{0.90}{\includegraphics{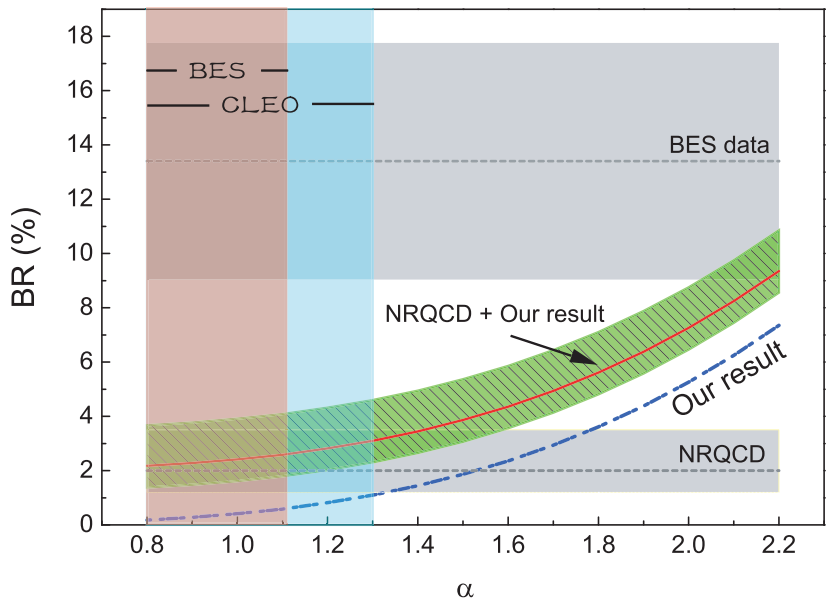}} \caption{The
comparison of our result (blue dash-dotted line) with the BES data
(dashed line with shadow band) of excessive non-$D\bar D$
component of the inclusive $\psi(3770)$ decay
\cite{Ablikim:2007zz} and the result of $\psi(3770)\to
\mathrm{Light\, Hadrons}$ (dash-dotted line with shadowed band) by
the color-octet mechanism calculated up to next to leading order
within the framework of NRQCD \cite{He:2008xb}. Here the red line
with green shadowed band is the total result including the NLO
NRQCD effects and FSI contribution. The green shadowed band
corresponds to the error tolerance, coming from the NRQCD estimate
in Ref. \cite{He:2008xb}. The orange and light blue shadowed bands
are the suitable window for $\alpha$, which is respectively
determined by the BES data  \cite{Ablikim:2005cd} and CELO data
\cite{Adams:2005ks} of $\psi(3770)\to \rho\pi$. \label{compare}}
\end{figure}

The branching ratio of $\psi(3770)\to \mathrm{non}-D\bar D$
including all $J/\psi\eta,\,\rho\pi,\,\omega\eta,\,K^*K$ modes
with a fixed value $\mathcal{I}=0.4$ GeV$^{-2}$ is shown in Fig.
\ref{compare} within the range of $\alpha=0.8\sim2.2$.
Furthermore, let us compare our result with the BES data
\cite{Ablikim:2007zz} and the result of $\psi(3770)\to
\mathrm{Light\, Hadrons}$ including the color-octet mechanism
calculated up to next to leading order in the approach of NRQCD
\cite{He:2008xb}. Fig. \ref{compare} shows that when FSI effects
are taken into account, the NRQCD results plus FSI contribution
can be very close to the BES data as long as $\alpha$ takes a
value of 2.0$\sim$2.2. Since our results heavily depend on the
parameter $\alpha$, which is fully determined by the
non-perturbative QCD effects and therefore cannot be determined
based on a first principle, one can only phenomenologically fix it
by fitting data. We also notice that the amplitudes in eqs.
(\ref{aaaa1})-(\ref{aaaa2}) are dependent on the values of
coupling constant in every vertex, which results in that the decay
width is proportional to the square product of all of the coupling
constants. If the uncertainty is $20\%$ for each coupling
constant, the maximum of uncertainty the decay width is $4\%$.

\begin{center}
 \begin{ruledtabular}
  \begin{table}
    \begin{tabular}{c|ccccc}
$\alpha$&$\Gamma_{\rho\pi}$
(keV)&$\Gamma_{\mathrm{non}-D\bar{D}}^{FSI}\,(\mathrm{keV})$&
$\mathcal{B}_{\mathrm{non}-D\bar{D}}^{FSI}\,(\%)$&$\mathcal{B}[NRQCD+Ours]\,(\%)$\\\hline
0.8&20&48&0.2&$1.4\sim3.7(2.2)$\\
0.9&32&75&0.3&$1.5\sim3.8(2.3)$\\
1.0&47&113&0.4&$1.6\sim3.9(2.4)$\\
1.1&66&160&0.6&$1.8\sim4.1(2.5)$
\\1.2&94&223&0.8&$2.0\sim4.3(2.8)$
\\1.3&127&301&1.1&$2.3\sim4.6(3.1)$
    \end{tabular}
    \caption{The typical values of decay width of $\rho\pi$
    channel $\Gamma_{\rho\pi}$, the sum of decay widths $\Gamma_{\mathrm{non}-D\bar{D}}^{FSI}$ and the branching
    ratio $\mathcal{B}_{\mathrm{non}-D\bar{D}}^{FSI}$
    of all channels discussed in this work. Here we fix $I=0.4$
    GeV$^{-2}$. The branching fraction $\mathcal{B}[NRQCD+Ours]$, which is the sum of our result and the NRQCD
    result. The values in bracket are the central values.
    \label{values}}
    \end{table}
 \end{ruledtabular}
\end{center}

The BES data $\mathcal{B}[\psi(3770)\to \rho\pi]<2.4\times10^{-3}$
with corresponding width $\Gamma[\psi(3770)\to \rho\pi]<65$ keV
and the CLEO data $\mathcal{B}[\psi(3770)\to
\rho\pi]<4.0\times10^{-3}$ with corresponding width
$\Gamma[\psi(3770)\to \rho\pi]<109$ keV \cite{Adams:2005ks} help
to further constrain the range of $\alpha$ to $0.8<\alpha<1.1$ and
$0.8<\alpha<1.3$, respectively. The relevant values of the decay
widths and the branching fraction are listed in Table
\ref{values}. It is noted that when the FSI is taken into account
and $\alpha$ is much restricted, the prediction of the branching
ratio of $\psi(3770)\to \mathrm{non}-D\bar{D}$ caused by the FSI
can reach up to
$\mathcal{B}_{\mathrm{non}-D\bar{D}}^{FSI}=(0.2\sim1.1)\%$ (taking
CLEO data of $\psi(3770)\to \rho\pi$ to constrain $\alpha$). It
indicates that even though FSI is significant, it cannot make a
drastic change as long as $\alpha$ is restricted to be less than
2.1. Furthermore, the upper limit of the total contribution of the
NRQCD and FSI is up to $4.6\%$. The branching ratios of E1
transition $\psi(3770)\to \gamma \chi_{cJ}$ (J=0,1,2) and
$\psi(3770)\to J/\psi\pi\pi,\,J/\psi\eta$ are about $(1.5\sim
1.8)\%$ \cite{Amsler:2008zz,Coan:2005ps,Briere:2006ff}. If summing
up all the above non-$D\bar{D}$ contributions, the branching ratio
of the channels with non-$D\bar{D}$ final states  can be as large
as $6.4\%$, which is still smaller than the experimental value
$\mathcal{B}[\psi(3770)\to \mathrm{non}-D\bar{D}]=(13.4\pm5.0\pm
3.6)\%$ but near its lower bound \cite{Ablikim:2007zz}.

As a short summary, let us emphasize a few points. First, even
including contributions of color-octet, the NRQCD prediction on
the branching ratio of $\psi(3770)\to \mathrm{non}-D\bar D$ which
is calculated up to NLO, cannot coincide with the data of BES
\cite{He:2008xb}. At the energy range, the FSI obviously is
significant and this allegation has been confirmed by many earlier
phenomenological studies on other processes. When the FSI effects
are taken into account, the discrepancy between theoretical
prediction and data is significantly alleviated, even though not
sufficient. Considering the rather large error range in
measurements of both inclusive $\mathrm{non}-D\bar D$ decay of
$\psi(3770)$ and the exclusive mode $\psi(3770)\to \rho\pi$, one
would still be able to obtain a value for the parameters $\alpha$
which does not conflict with the data, by which the theorical
prediction and data might be consistent. The more accurate
mesurements which will be conducted in the future will provide
more information which can help to make a definite conclusion if
the FSI indeed solves the "puzzle" or not. Secondly, our result
shows that the FSI can make significant contribution to all the
channels of $\psi(3770)\to \rho\pi,\,K^*K$, and each of them
should be searched in future experiments. Thirdly, no doubt, more
accurate measurements on $\psi(3770)\to \mathrm{non}-D\bar{D}$,
especially $\psi(3770)\to \mathrm{Light\, Hadrons}$, are
necessary. Thanks to the great improvement of facility and
technology of detection at the charm-tau energy region, the BESIII
\cite{Asner:2008nq} will provide much more precise data, by which
we may gain more information. Furthermore, along the other lines
more theoretical studies which may involve other mechanics, even
new physics beyond standard model are badly needed.

\vspace{0.5cm}

\noindent {\bf Acknowledgement} We thank Prof. Kuang-Ta Chao,
Prof. Shi-Lin Zhu, Prof. Eef van Beveren for fruitful discussions.
We also would like to thank Prof. Chang-Zheng Yuan from BES
collaboration, Dr. Brian Heltsley and Dr. Hajime Muramatsu from
CLEO collaboration for reminding us of experimental results. We
acknowledge the National Natural Science Foundation of China and
the Special Grant for Ph.D programs of the Education Ministry of
China. One of us (X.L.) is also partly support by the
\emph{Funda\c{c}\~{a}o para a Ci\^{e}ncia e a Tecnologia} \/of the
\emph{Minist\'{e}rio da Ci\^{e}ncia, Tecnologia e Ensino Superior}
\/of Portugal, under contract SFRH/BPD/34819/2007.

\end{document}